\documentclass[reprint,
superscriptaddress,
 amsmath,amssymb,
 aps
]{revtex4-1}
\usepackage{xcolor}
\usepackage{graphicx}
\usepackage{dcolumn}
\usepackage{bm}
\usepackage{float}
\usepackage[normalem]{ulem}

\newcommand{\rev}[1]{\textcolor{black}{#1}}

\begin{document}

\preprint{APS/123-QED}

\title{Nonlinear magnon polaritons}

\author{Oscar Lee}
 \affiliation{London Centre for Nanotechnology, University College London, London WC1H 0AH, United Kingdom}
 
\author{Kei Yamamoto}
 \email{yamamoto.kei@jaea.go.jp}
\affiliation{Advanced Science Research Center, Japan Atomic Energy Agency, 2-4 Shirakata, Tokai 319-1195, Japan}

\author{Maki Umeda}
\affiliation{Advanced Science Research Center, Japan Atomic Energy Agency, 2-4 Shirakata, Tokai 319-1195, Japan}
 \affiliation{Institute for Materials Research, Tohoku University, Sendai 980-8577, Japan}
 
\author{Christoph W. Zollitsch}
 \affiliation{London Centre for Nanotechnology, University College London, London WC1H 0AH, United Kingdom}

 \author{Mehrdad Elyasi}
 \affiliation{WPI Advanced Institude for Materials Research, Tohoku University, 2-1-1, Katahira, Sendai 980-8577, Japan}
 
\author{Takashi Kikkawa}
 \affiliation{Department of Applied Physics, The University of Tokyo, Tokyo 113-8656, Japan}

\author{Eiji Saitoh}
\affiliation{Advanced Science Research Center, Japan Atomic Energy Agency, 2-4 Shirakata, Tokai 319-1195, Japan}
\affiliation{WPI Advanced Institude for Materials Research, Tohoku University, 2-1-1, Katahira, Sendai 980-8577, Japan}
\affiliation{Department of Applied Physics, The University of Tokyo, Tokyo 113-8656, Japan}
\affiliation{Institute for AI and Beyond, The University of Tokyo, Tokyo 113-8656, Japan.}

\author{Gerrit E. W. Bauer}
 \affiliation{WPI Advanced Institude for Materials Research, Tohoku University, 2-1-1, Katahira, Sendai 980-8577, Japan}
 
 \author{Hidekazu Kurebayashi}
 \email{h.kurebayashi@ucl.ac.uk}
 \affiliation{London Centre for Nanotechnology, University College London, London WC1H 0AH, United Kingdom}

\begin{abstract}
We experimentally and theoretically demonstrate that nonlinear spin-wave
interactions suppress the hybrid magnon-photon quasiparticle or
\textquotedblleft magnon polariton\textquotedblright\ in microwave spectra
of an yttrium iron garnet film detected by an on-chip split-ring resonator.
We observe a strong coupling between the Kittel and microwave cavity
mode\rev{s} in terms of an \rev{avoided crossing} as a function of magnetic fields at low
microwave input powers, but a complete closing of the gap at
high powers. The experimental results are well explained by a theoretical
model including the three-magnon decay of the Kittel magnon into spin waves.
The gap closure originates from the saturation of the ferromagnetic
resonance above the Suhl instability threshold by a coherent back reaction from the spin waves.
\end{abstract}

\maketitle

\preprint{APS/123-QED}

\affiliation{London Centre for Nanotechnology, University College London,
London WC1H 0AH, United Kingdom}

\affiliation{Advanced Science Research Center, Japan Atomic Energy Agency,
2-4 Shirakata, Tokai 319-1195, Japan}

\affiliation{Advanced Science Research Center, Japan Atomic Energy Agency,
2-4 Shirakata, Tokai 319-1195, Japan} 
\affiliation{Department of Applied
Physics, Tohoku University, Aoba 6-6-05, Sendai, 980-8579, Japan}

\affiliation{
 London Centre for Nanotechnology, University College London, London WC1H 0AH, United Kingdom}

\affiliation{WPI Advanced Institude for Materials Research, Tohoku
University, 2-1-1, Katahira, Sendai 980-8577, Japan}

\affiliation{Department of Applied Physics, The University of Tokyo, Tokyo
113-8656, Japan}

\affiliation{INNOVENT e.V. Technologieentwicklung, 07745 Jena, Germany}

\affiliation{WPI Advanced Institute for Materials Research, Tohoku
University, 2-1-1, Katahira, Sendai 980-8577, Japan} 
\affiliation{Department
of Applied Physics, The University of Tokyo, Tokyo 113-8656, Japan}

\affiliation{WPI Advanced Institute for Materials Research, Tohoku
University, 2-1-1, Katahira, Sendai 980-8577, Japan}

\affiliation{London Centre for Nanotechnology, University College London,
London WC1H 0AH, United Kingdom}

\rev{The spectral properties of many-body systems can often be understood in terms of weakly interacting quasiparticles. When tuning the energies of two elementary excitations into degeneracy by an external parameter, their coupling leads to a level repulsion.  When the resultant gap is larger than the level broadening, it becomes observable in the spectrum. This so-called strong coupling generates a hybrid quasiparticle that shares the properties of both ingredients. The strong coupling between magnons, phonons, photons, excitons, plasmons, \emph{etc}. has important consequences and applications in condensed matter physics~\cite{Haroche_book2006,Schoelkopf_Nature2008,Dovzhenko_Review2018,Diaz_RMP2019,Basov_Science2016,Bozhko_LTP2020}. Here we address the magnon polariton, \emph{i.e.} the mixed state of a spin wave in a ferromagnet and a microwave magnetic field~\cite{Rameshti_arXiv2021,Harder_JAP2021,Quirion_APEX2019,Awschalom_IEEETQE2021}. While magnon polaritons are often discussed in the context of quantum computing by discrete qubits~\cite{Tabuchi_PRL2014}, they are more generally relevant for the control of continuous magnon variables by electromagnetic fields. Although they have been extensively studied in the linear response regime of weak microwave excitation, their nonlinearities have so far escaped similar attention.}

\rev{In comparison, nonlinearites in magnetic excitations have been known for many decades~\cite{Bloembergen_PR1952,L'vov_book1994,Wigen_book1994}. They can be useful in, for instance, probabilistic bits~\cite{Makiuchi_APL2021,Elyasi_PRB2022}, and offer continuous variables with controllable squeezing and entanglement that act as resources in quantum information~\cite{YUAN_PhysRep2022,Elyasi_PRB2020}. The magnon nonlinearities can be treated systematically by the Holstein-Primakoff power expansion of a spin Hamiltonian in creation and annihilation operators~$b_{\bm{k}}^{\dagger },b_{\bm{k}}$. With increasing excitation, progressively higher-order terms of the expansion become important. We focus on the three-magnon scattering; the leading nonlinear term that involves the splitting of a magnon into two and the reciprocal confluence~\cite{Suhl_1957,L'vov_book1994}. The interaction causes the first-order Suhl instability of a uniform precession of the magnetic order, or Kittel mode represented by~$b_0$, at a threshold power that can be very small in low-damping magnets~\cite{Suhl_1957,Synogach_PRL2000,Mathieu_PRB2003,Romero_PRB2009,Schultheiss_PRL2009,Kurebayashi_NatMat2011,Sakimura_NComm2014,Barsukov_SciAdv2019}. Figure~1(a) illustrates the scattering process in which a Kittel magnon decays into two magnons of half its frequency and opposite momenta~$\pm \bm{k}$. This three-magnon splitting is allowed only when magnetic dipole-dipole interactions render a nonmonotonic magnonic dispersion with minima at half the Kittel mode frequency or below. When the Kittel mode is excited, the three-magnon splitting pumps the magnon pair amplitude~$\langle b_{ \bm{k}}b_{-\bm{k}}\rangle $ at a rate proportional to that of the Kittel mode~$|\langle b_0 \rangle |$. When the pumping rate exceeds the relaxation rate of the magnons~$\eta _{\bm{k}}$, a nonthermal magnon population accumulates in the valleys of the magnon dispersion. This first-order Suhl instability manifests itself in microwave reflection spectra by \emph{e.g.} distortions of the spectral line shape from a Lorentzian~\cite{L'vov_book1994}.}

\begin{figure}[t]
\centering
\includegraphics[scale=0.29]{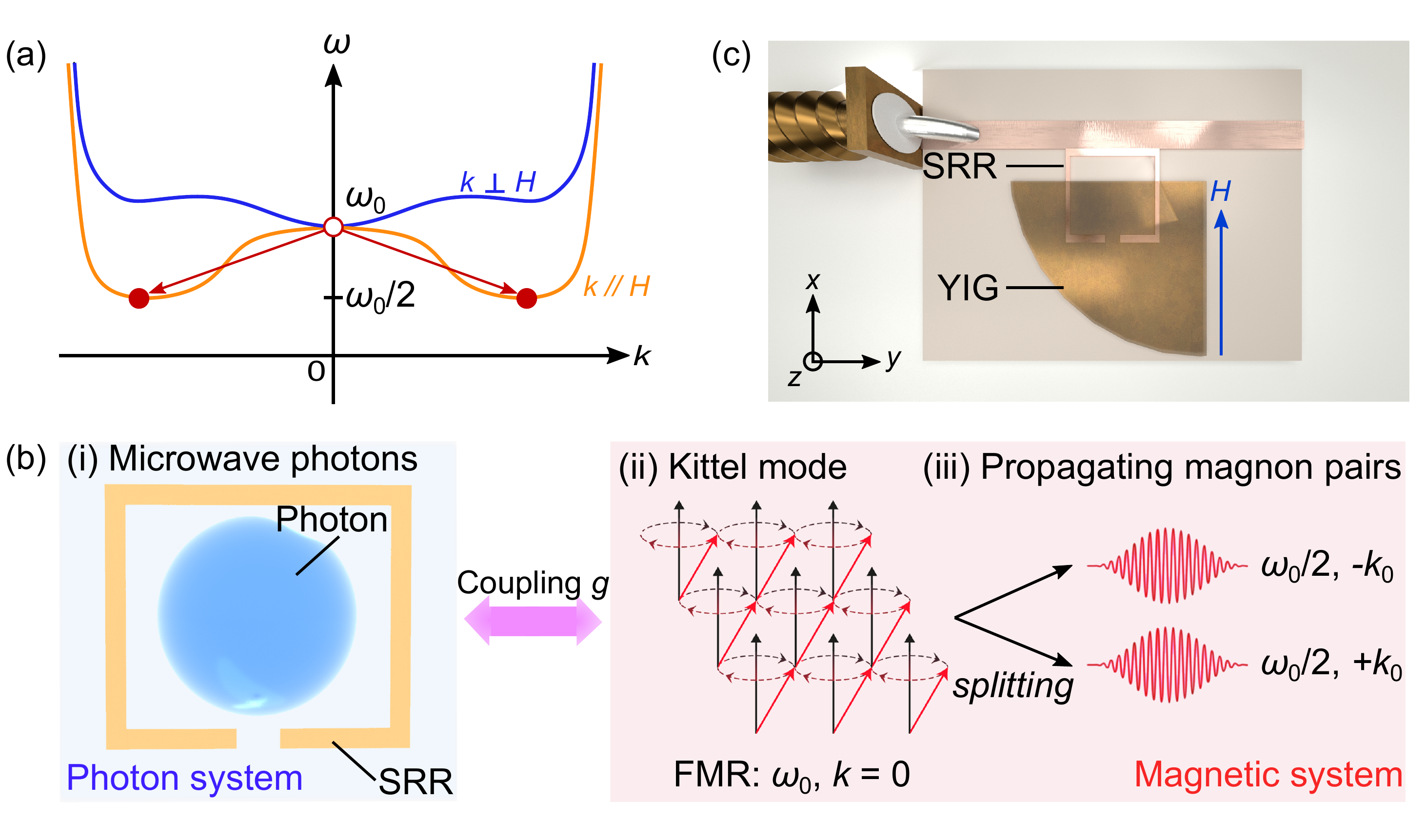}
\caption{\rev{(a) Spin-wave dispersion of a YIG film. Red arrows indicate the split of a Kittel magnon with the frequency~$%
\protect\omega _{\text{0}}$ into two magnons of frequency~$\protect\omega _{%
\text{0}}/2$. (b) Schematic of the magnon polariton with the spin-wave instability as explained in the main text. (c) Measurement set-up.}}
\label{fig:Fig_1}
\end{figure}

\rev{In this Letter, we study this nonlinear instability under the condition that the Kittel magnon strongly couples to the photon in a discrete microwave cavity~[Fig.~\ref{fig:Fig_1}(b)]. Since magnon-photon coupling can be used to read information out or distill entanglement in these applications, nonlinear magnon-polariton phenomena may become a crucial ingredient in novel computing and information-technology paradigms~\cite{YUAN_PhysRep2022,Rameshti_arXiv2021,Elyasi_PRB2022}. Our main result is the observation and modeling of the suppression of the strong-coupling gap by the instability. The nonlinear spin-wave equation coupled to the cavity mode explains our observations in terms of the saturation of the Kittel mode by a dynamical phase correlation between the cavity photons and the magnon pairs in the valleys. To the best of our knowledge, a tunable strong coupling has not been reported for magnon polaritons and adds to the appeal of magnetic materials for classical and quantum information technologies~\cite{Mahmoud_JAP2020,YUAN_PhysRep2022,Rameshti_arXiv2021}.}

\begin{figure*}[tbp]
\centering
\includegraphics[scale=0.6]{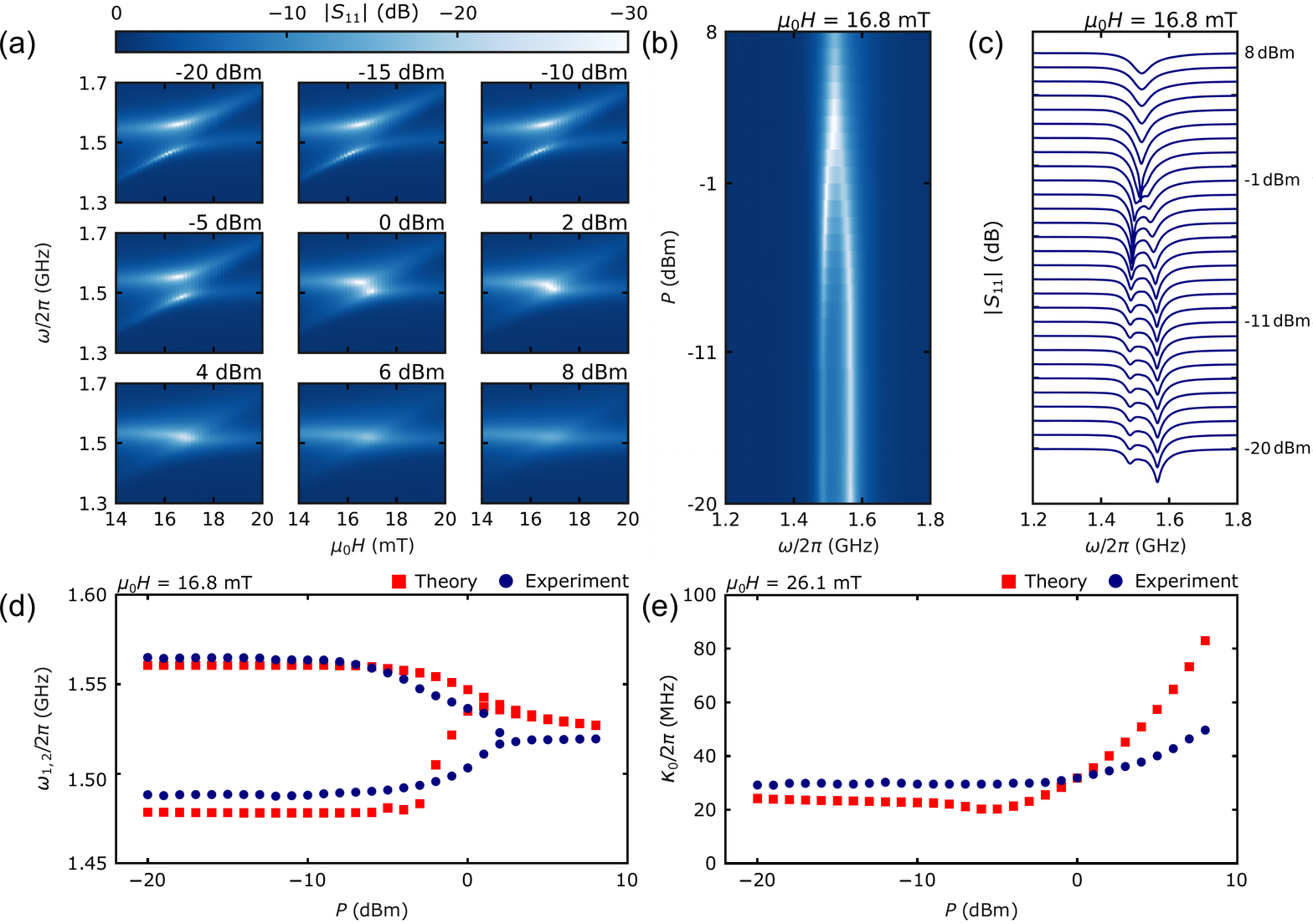}
\caption{(a) Microwave absorption spectra (\rev{$|S_{11}|$}~(dB)) as a function of
microwave frequency and magnetic field strength, for different microwave power ranging from -20 to 8~dBm.
(b)\rev{\textendash}(c) Collection of frequency domain scans for a fixed magnetic field of
16.8~mT. \rev{(d) Comparison between measured and calculated gaps.} Peak \rev{frequencies} are extracted from individual fits for different microwave powers. \rev{(e)} Microwave power evolution of \rev{the observed and calculated linewidths of the Kittel mode in the weak coupling regime ($\mu _0 H=$ 26.1~mT).}}
\label{fig:Fig_2}
\end{figure*}

\begin{figure*}[tbp]
\centering
\includegraphics[scale=0.6]{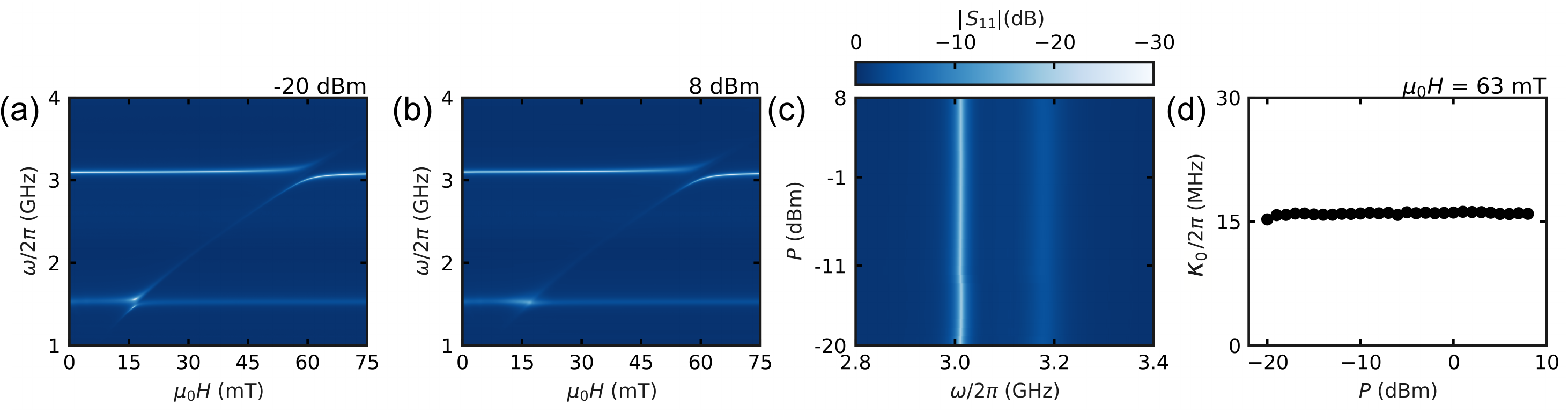}
\caption{$|S_{11}|$ as a function of microwave frequency and magnetic field strength, for low (a) and high
(b) microwave powers. \rev{(c)} Microwave absorption spectra for the 3~GHz SRR mode at a fixed magnetic
field of 60~mT, \rev{at which the frequency difference between the two peaks is the smallest}. (d) Power
evolution of Kittel mode linewidth (\rev{$\kappa_0$}) at a
fixed magnetic field above the \rev{avoided crossing} (63~mT) for the 3~GHz SRR mode.}
\label{fig:Fig_3}
\end{figure*}

We study a 5~$\mathrm{\mu m}$~thick YIG film grown on a gadolinium
gallium garnet substrate by liquid-phase-epitaxy~\cite{Dubs_2017, Dubs_PRM2020} and placed on a split-ring microwave
resonator~(SRR) as depicted in Fig.~\ref{fig:Fig_1}(c)~\cite{Stenning_OptExp2013,Bhoi_JAP2014}. We measured
microwave absorption/reflection spectra \rev{$|S_{11}|$} at room temperature using a
vector network analyzer~(VNA). Figure~\ref{fig:Fig_2}(a) shows \rev{$|S_{11}|$} 
for different input microwave
powers~$P$ as a function of frequency~$\omega \rev{/2\pi } $ and magnetic field~$\rev{\mu _0}H$. For $P=-20$~dBm, a prominent \rev{avoided crossing} between the \rev{Kittel mode} frequency~$\omega _0 \rev{/2\pi }$ and the cavity photon mode~$\omega _{r}\rev{/2\pi}\approx 1.5$~GHz is evidence for strong
magnon-photon coupling. \rev{The minimum frequency difference, half of which is the coupling strength \protect{$g/2\pi=41~\mathrm{MHz}$}, occurs at the resonance field $\mu _0 H_{\rm res}=16.8$~mT. In linear response,} $\left\vert g\right\vert =\eta \gamma \sqrt{\hbar
\mu _{0}\omega _{r}}\sqrt{N/\mathcal{V}_{c}}$,
where $\eta $, $\gamma $, $\hbar $, $\mu _{0}$, $N$, and $\mathcal{V}_{c}$
are the filling factor that characterizes the spatial mode overlap between the photon and magnon modes, the gyromagnetic ratio, reduced Planck constant, vacuum permeability, number of spins, and the cavity mode volume, respectively~\cite{Rameshti_arXiv2021}. \rev{The individual linewidths of the photon and magnon are obtained by Lorentzian function fittings of the respective resonances~$\kappa _{\rm r}(\kappa _0 )/2\pi=52.0(28.0)~\mathrm{MHz}$ at $\mu _0 H =26.1$~mT far from the avoided crossing.} 
With increasing $P$, the avoided crossing gap narrows and the
two peaks eventually merge~[Fig.~\ref{fig:Fig_2}(a)]. \rev{Figures~\ref{fig:Fig_2}(b) and (c) show that the two peaks on resonance $H=H_{\rm res}$ coalesce into a single one at high powers, seemingly cancelling the magnon-photon coupling}. Figure~\ref{fig:Fig_2}(d) summarizes \rev{the frequencies of the peaks in the spectra, illustrating the vanishing of the gap that constitutes our main result. As argued in the following, we attribute it to the first-order Suhl instability.}

The Suhl instability alters magnetic susceptibility~\cite{Damon_RMP1953,Bloembergen_PR1954,Suhl_1957} and lineshape~\cite{L'vov_book1994} \rev{by the} nonlinear back reaction of the \rev{excited magnon pairs on the Kittel magnon. We confirm an implied increased broadening by measuring the $P$ dependence of the Kittel mode linewidth at $\mu _0 H =26.1~\textrm{mT}>\mu _0 H_{\rm res}$, far away from the cavity resonance. As shown in Fig.~\ref{fig:Fig_2}\rev{(e)}, we observe an increase in broadening for $P \gtrsim 0 $~dBm, which is expected for entering the power regime of the first-order Suhl instability. The critical number of Kittel magnons per spin at the threshold is $|\langle b_0 \rangle |^2 /N = {\rm const} \times \eta _{\bm{k}}^2 /\omega _M^2$ where $\omega _M = \gamma \mu _0 M_s$ is the saturation magnetization. The difference between the onset powers for the gap closure in Fig.~\ref{fig:Fig_2}(d) and the broadening in Fig.~\ref{fig:Fig_2}(e) implies that the dimensionless constant of order unity depends on the system parameters including $H$ and $\omega $.}  
\begin{figure*}[tbp]
\includegraphics[scale=0.60]{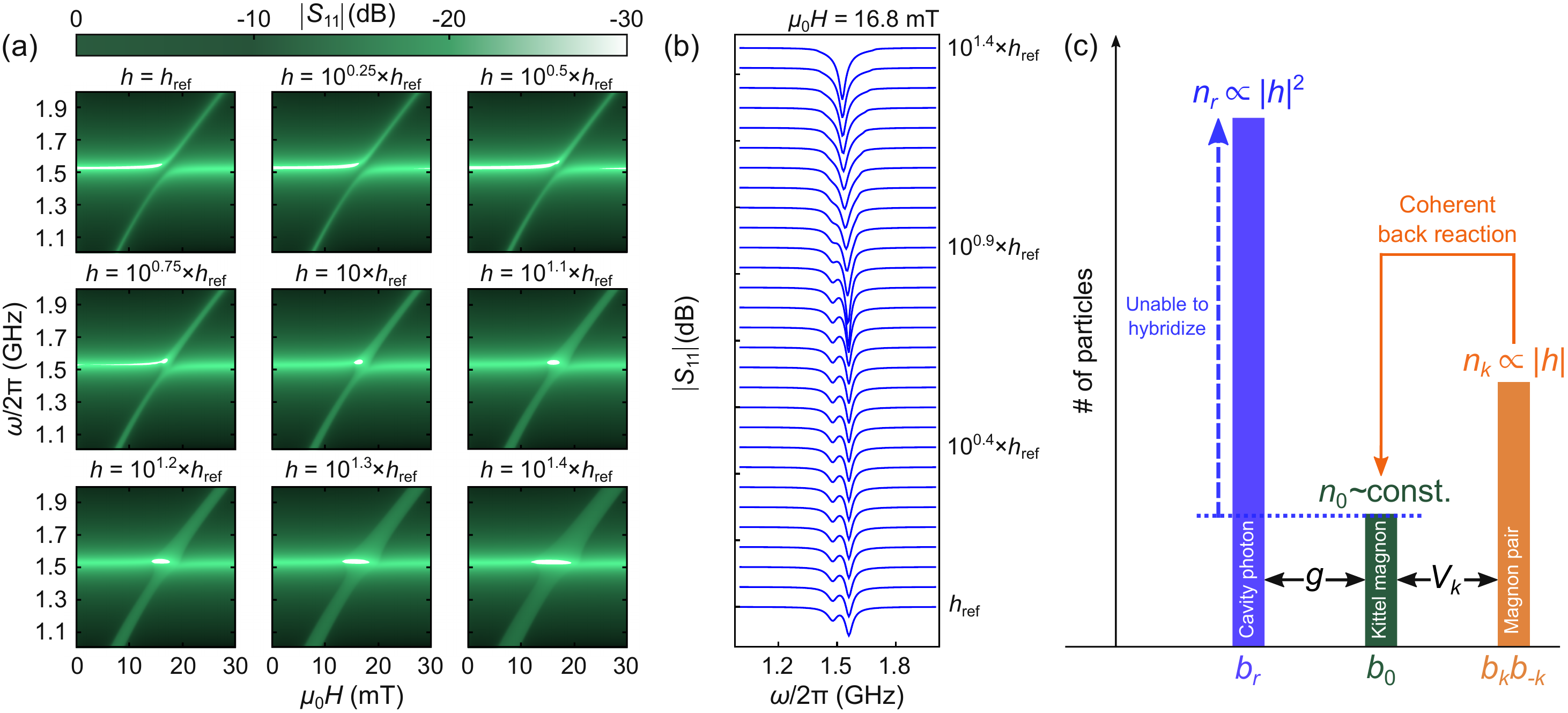}
\caption{\rev{(a)} Microwave reflection spectrum \rev{$|S_{11}|$} (defined in Supplementary Materials) calculated from
Eqs.~(\protect\ref{eq:saturation}) and (\protect\ref{eq:cavity}) for model
parameters in the text and an arbitrary reference input power $h_{\mathrm{ref%
}}$. \rev{(b)} \rev{$|S_{11}|$} at $\mu _{0}H=16.8$~mT. \rev{(c)} Schematic of the particle number growth with $h$ for magnon polaritons with the Suhl instability\rev{.}}
\label{fig:Fig_4}
\end{figure*}

\rev{We can corroborate our interpretation by increasing $H$ to couple the Kittel mode with a higher SRR cavity mode. Energy conservation $\omega _0 = 2\omega _{\bm{k}}$, where $\omega _{\bm{k}}$ is the frequency of a magnon with wavevector $\bm{k}$, demands that $\omega _0 \geq 2\omega _{\rm b}$ where $\omega _{\rm b}$ is the band edge frequency. Since both $\omega _0 $ and $\omega _{\rm b}$ increase roughly linearly with $H$, the three-magnon splitting is forbidden above a critical field value~$H_{\rm cr}$, at which} $\omega _{0}\left( H_{\mathrm{cr}}\right)\rev{/2\pi}=2.\rev{59}\,\mathrm{%
GHz}$ for our YIG sample with a thickness of 5~$\mathrm{\mu m}$, $%
M_{\rm s}=1.\rev{26}\times 10^{5}$~A/m, and a stiffness constant of $\lambda _{\rm ex}=3\times 10^{-16}$~m$^2$~\cite{Stancil_book2008} (see SM~\footnote{See Sec. II of Supplementary Material for the basic characteristics of spin waves in a YIG film}). The magnon polariton of the 3~GHz SRR mode in Fig.~\ref{fig:Fig_3}(a) should therefore depend much less on the microwave power. By increasing $P$ up to 8~dBm as before, the reflection spectrum [Fig.~\ref{fig:Fig_3}(b)] and the fixed-field plot in Fig.~\ref{fig:Fig_3}(c) confirm
that the \rev{avoided crossing} gap does not vanish and \rev{Kittel mode linewidth in Fig.~\ref{fig:Fig_3}(d) remains unchanged}, which supports our hypothesis that
the Suhl instability explains Fig.~\ref{fig:Fig_2}.

\rev{We substantiate the above arguments by the kinetic theory of nonlinear spin wave dynamics~\cite%
{L'vov_book1994,Suhl_1988} extended to incorporate the magnon polariton. We start from the model Hamiltonian~$%
\mathcal{H}=\mathcal{H}_1 + \mathcal{H}_{2}+\mathcal{H}_{3}$ (in frequency units), in which  
\begin{equation}
    \mathcal{H}_2 = \omega _r b_r^{\dagger }b_r +\left[ gb_0^{\dagger }b_r +{\rm h.c.} \right]  + \omega _0 b_0^{\dagger }b_0 + \sum _{\bm{k}\neq \bm{0}}\omega _{\bm{k}}b_{\bm{k}}^{\dagger }b_{\bm{k}}
\end{equation}
describes non-interacting fields as coupled harmonic oscillators, where $b_r$ is the annihilation operator for the selected cavity photon, and $\omega _{r}$ is \rev{its} frequency. The microwave stripline drive contributes
\begin{equation}
    \mathcal{H}_1 = \left[ h e^{-i\omega t} \left( U_r b_r^{\dagger }+U_0 b_0^{\dagger } \right) +{\rm h.c.} \right] ,
\end{equation}
where $h$ and $\omega $ are the amplitude (in frequency units) and frequency of the stripline field, and $U_{0}$ and $U_{r}$ are its
(dimensionless) coupling strengths to the Kittel and cavity mode,
respectively. The nonlinear coupling $V_{\bm{k}}\sim \omega _M /\sqrt{N}$ in 
\begin{equation}
\mathcal{H}_{3}=\frac{1}{2}\sum_{\bm{k}\neq \bm{0}}\overline{V_{\bm{k}}}b_{0}b_{\bm{k}%
}^{\dagger }b_{-\bm{k}}^{\dagger }+\mathrm{h.c.}
\end{equation}%
is a function of the material parameters~\cite{L'vov_book1994}. Overlines denote complex conjugation throughout.}
We omitted four-magnon scattering terms because in the present setup the critical power of the first order Suhl instability is much smaller than the second order one (see SM~\footnote{See Sec. IV of Supplementary Material for the theoretical details of the Suhl instability}). \rev{At room temperature, one may safely interpret the field operators as classical amplitudes with thermal fluctuations.}
In the film geometry with an in-plane static magnetic field, only a narrow band of magnons are involved in the onset of the instabilities~\cite{Zakharov_1975}, which we approximate here by the single pair $\pm \bm{k} \parallel \bm{H}$ with smallest $\eta _{\bm{k}}/\left| V_{\bm{k}} \right| $\rev{. The steady-state solutions are characterized by the thermal averages $\langle b_{0,r}\rangle $
and $\langle b_{\bm{k}}b_{-\bm{k}}\rangle $.} The coherent amplitude of the Kittel mode $\langle b_{0}\rangle $ is a root of a (complex) cubic algebraic equation (Eq.~(S25) in the SM), \rev{which at sufficiently high powers $|h| \rightarrow \infty $ approaches}
\begin{equation}
\langle b_{0}\rangle \rightarrow -{\rm e}^{-i\omega t+i\psi _{\bm{k}}} \overline{c_{\rm cr}} , \quad c_{\rm cr} = \frac{%
\omega _{\bm{k}}-\omega /2+ i\eta _{\bm{k}}}{V_{\bm{k}}} \label{eq:saturation}
\end{equation}%
where $\psi _{\bm{k}}$ is the phase of the magnon pair amplitude $\langle
b_{\bm{k}}b_{-\bm{k}}\rangle $. The absence of $h$ on the r.h.s. implies 
\emph{saturation}, i.e., the number of Kittel magnons \rev{$n_0$ cannot exceed} the critical value $| c_{\mathrm{cr}} | ^2$,
which depends only on the \rev{magnonic parameters}. Furthermore \rev{(for all $|h|$)}, 
\begin{align}
\langle b_{r}\rangle =& \frac{\overline{g}\langle b_{0}\rangle
+hU_{r}{\rm e}^{-i\omega t}}{\omega -\omega _{r}+i\rev{\kappa _{r}}},  \label{eq:cavity} \\
\langle b_{\bm{k}}b_{-\bm{k}}\rangle =& -\frac{c_{\mathrm{cr}}\langle
b_{0}\rangle }{|c_{\mathrm{cr}}|^{2}-|\langle b_{0}\rangle |^{2}}\frac{k_{B}T%
}{\hbar \omega _{\bm{k}}},
\end{align}%
where $T$ is the temperature and $k_{B}$ the Boltzmann constant. Photon and magnon pair amplitudes \rev{coherently oscillate with} the Kittel mode, \rev{whose phase in turn locks to that} of the driving field $h$. \rev{The saturation limit Eq.~(\ref{eq:saturation}) is valid in a nonvanishing interval above the critical power~\cite{Zakharov_1975} and explains the main features of the observations.}

Figures~\ref{fig:Fig_4}\rev{(a) and \ref{fig:Fig_4}(b)} summarize the theoretical results with the standard parameters for YIG, \rev{\emph{i.e.} an extracted saturation magnetization $M_{\rm s}=1.26\times 10^{5}$~A/m for $\gamma /2\pi =28$~GHz/T, and
a magnetic-relaxation-rate parameter $\kappa _{0}/2\pi=22$~MHz. 
We model the microwaves system by $\omega _{r}/2\pi =1.53$~GHz, $\kappa _r/2\pi=52$~MHz, $U_{r}=0.95\times e^{0.4i\pi }$, $U_{0}=0.31$. We take
magnon-photon coupling $g/2\pi =41$~MHz directly from the gap of the avoided crossing} at low $P$%
. The Kittel formula is $\omega _{0}=\mu _{0}\gamma \sqrt{H(H+M_{\rm s})}.$ For
the coherently coupled magnon pair at $\omega _{\bm{k}}=\omega /2$, we
assume $\eta _{\bm{k}}=0.01\times \omega _0 /2$ and $V_{\bm{k}}=\rev{\omega _M}\times 10^{-11/2}$ \rev{(see SM~\footnote{See Sec. IV of Supplementary Material for the rational behind the choice of the parameters}). The calculated spectra compare favorably with the \rev{observed} gap closure \rev{and} the lineshapes~[Figs.~\ref{fig:Fig_2}(d) and (e)]. Note that the spectra at high powers cannot be explained by the dissipative coupling observed in very different regimes in Refs.~\cite{Harder_PRL2018,Boventer_PRR2020}}. We can \rev{instead attribute} the quenching of the
\rev{avoided crossing to the saturation of the Kittel mode [Eq.~(\ref{eq:saturation})].}
Below the critical power, a photon injected into the cavity mixing with \rev{a} Kittel \rev{magnon} causes the \rev{avoided crossing}. \rev{Above the critical power, however, cavity photons are much more numerous than the Kittel magnons limited by the coherent back reaction from the magnon pairs, as illustrated in Fig.~\ref{fig:Fig_4}(c). The excess photons become effectively decoupled and therefore do not show the avoided crossing. Their spectral characteristics overwhelm the gap opened by the few saturated magnons, thereby causing an apparent closure of the gap.}

In summary, we discovered \rev{suppression of} the
strong magnon-photon coupling in \rev{highly excited} microwave cavities \rev{at} the first-order Suhl instability.
\rev{T}he closure of the hybridization gap \rev{calculated with} a nonlinear spin-wave model
coupled to a microwave cavity photon mode \rev{agrees quantitatively with the observations}. This
effect is a result of the phase coherence between the photons and the entire
spin wave system \rev{that saturates the number of Kittel magnons under large microwave drives}. The ability to coherently excite or detect magnon pairs in
the low energy valleys \rev{not only contributes to studying and controlling quantum entanglement of magnons~\cite{YUAN_PhysRep2022,Elyasi_PRB2020}, but also} opens new avenues in magnonics, such as the microwave
spectroscopy of magnon Bose Einstein condensates~\cite{Demokritov_Nature2006}. The present work promises ample room for unexpected discoveries in nonlinear magnonics as an exciting research frontier.

KY is supported by JST PRESTO Grant Number JPMJPR2LB, Japan and JSPS KAKENHI (Nos.~19K21040 and 21K1388), GB by JSPS KAKENHI Grant No. 19H00645, and TK \rev{and ES} by JST CREST (JPMJCR20C1 and JPMJCR20T2), JSPS KAKENHI (JP19H05600 and JP20H02599), as well as the Institute for AI and Beyond of the University of Tokyo. We thank C. Dubs of INNOVENT e.V. Jena, Germany, for providing additional YIG films. For the purpose of open access, the author has applied a Creative Commons Attribution (CC BY) licence to any Author Accepted Manuscript version arising.

\bibliography{apssamp}

\end{document}


\author{Oscar Lee}
 \affiliation{London Centre for Nanotechnology, University College London, London WC1H 0AH, United Kingdom}
 
\author{Kei Yamamoto}
 \email{yamamoto.kei@jaea.go.jp}
\affiliation{Advanced Science Research Center, Japan Atomic Energy Agency, 2-4 Shirakata, Tokai 319-1195, Japan}

\author{Maki Umeda}
\affiliation{Advanced Science Research Center, Japan Atomic Energy Agency, 2-4 Shirakata, Tokai 319-1195, Japan}
 \affiliation{Institute for Materials Research, Tohoku University, Sendai 980-8577, Japan}
 
\author{Christoph W. Zollitsch}
 \affiliation{London Centre for Nanotechnology, University College London, London WC1H 0AH, United Kingdom}

 \author{Mehrdad Elyasi}
 \affiliation{WPI Advanced Institude for Materials Research, Tohoku University, 2-1-1, Katahira, Sendai 980-8577, Japan}
 
\author{Takashi Kikkawa}
 \affiliation{Department of Applied Physics, The University of Tokyo, Tokyo 113-8656, Japan}

\author{Eiji Saitoh}
\affiliation{Advanced Science Research Center, Japan Atomic Energy Agency, 2-4 Shirakata, Tokai 319-1195, Japan}
\affiliation{WPI Advanced Institude for Materials Research, Tohoku University, 2-1-1, Katahira, Sendai 980-8577, Japan}
\affiliation{Department of Applied Physics, The University of Tokyo, Tokyo 113-8656, Japan}
\affiliation{Institute for AI and Beyond, The University of Tokyo, Tokyo 113-8656, Japan.}

\author{Gerrit E. W. Bauer}
 \affiliation{WPI Advanced Institude for Materials Research, Tohoku University, 2-1-1, Katahira, Sendai 980-8577, Japan}
 
 \author{Hidekazu Kurebayashi}
 \email{h.kurebayashi@ucl.ac.uk}
 \affiliation{London Centre for Nanotechnology, University College London, London WC1H 0AH, United Kingdom}


\title{\large Supplementary Material for ``Nonlinear magnon polaritons"}

\maketitle

\section{Experimental geometry and mode profile of the resonator}

\begin{figure}[H]
\centering
\includegraphics[scale=0.7]{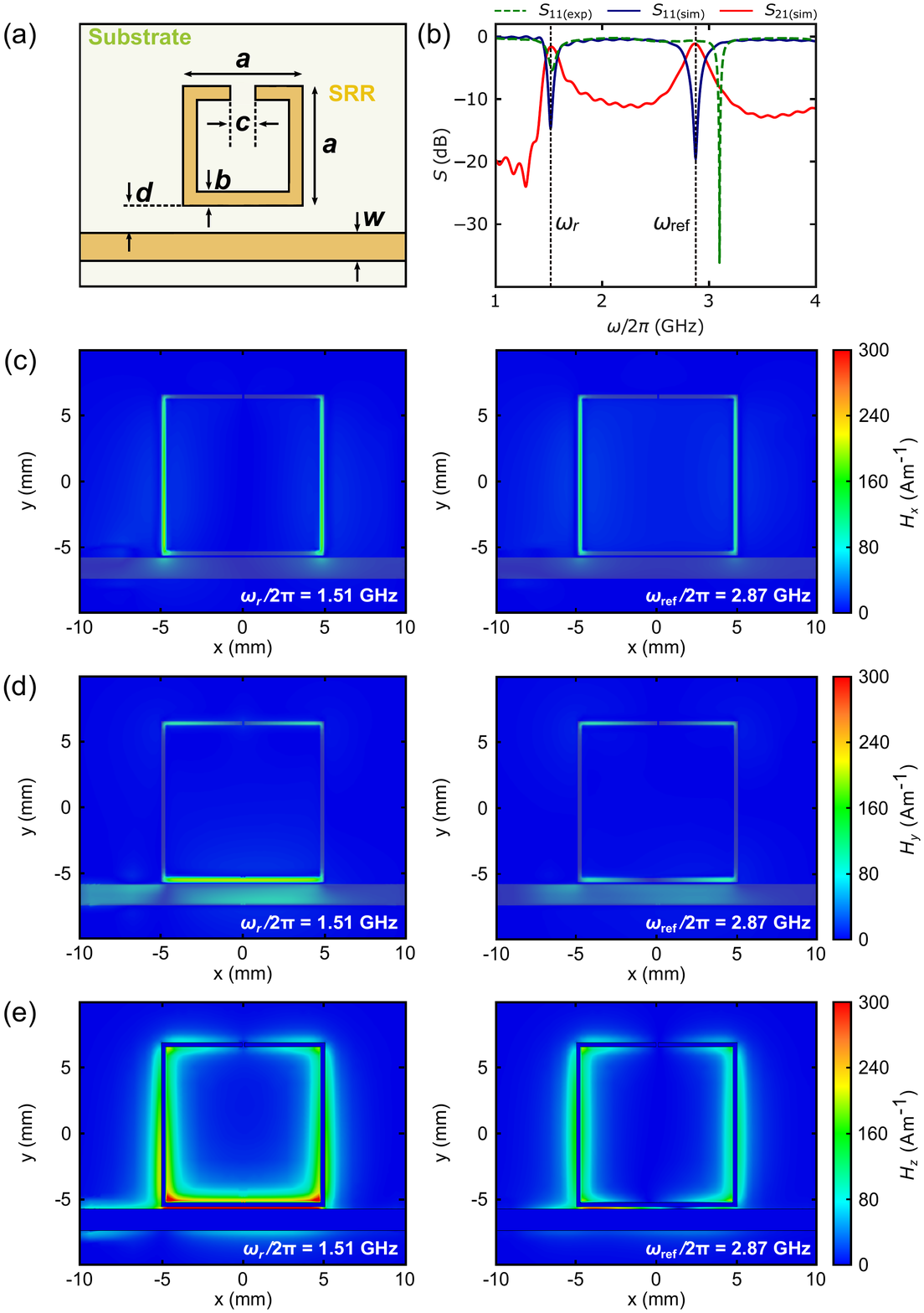}

    \caption{(a) Schematic drawing of the SRR used in the experimental setup. (b) Calculated microwave reflection and transmission spectra (\rev{$|S_{11}|$ and $|S_{21}|$}) as a function of frequency of dimensions depicted in (a). (c)-(e) Surface plots of local magnetic field distribution components for (c) $H_x$, (d) $H_y$ and (e) $H_z$ of the \rev{$|S_{11}|$} mode at the primary ($\omega_{r}/2\pi =1.51$~GHz) (left) and the secondary ($\omega_{\rm ref}/2\pi =2.87$~GHz) (right) mode frequencies, respectively.}
    
\label{fig:Fig_S1}
\end{figure}

We employed a combination of the time domain and the eigenmode solvers from the CST STUDIO SUITE 2021 package to simulate and calculate the response to a microwave input signal and its corresponding field distributions on the split ring resonator (SRR) used in our experiments. The schematic and dimensions of the SRR are shown in Fig.~\ref{fig:Fig_S1}(a). A square-shaped resonator that consists an outer length of $a = 10$~mm and an inner width of $b = 0.25$~mm with a gap of distance $c = 0.125$~mm has been prepared with a 17.5~$\mu$m double-sided copper coating, fabricated on a ROGERS RT6010LM substrate of height~$h = 1.5$~mm with a relative permittivity~$\epsilon_r=10.7$ and loss~$\tan\delta=2.3\times10^{-3}$. A 50~$\ohm$ impedance-matched microstrip line of width~$w = 1.5$~mm is inductively coupled to the SRR, separated by a distance~$d = 0.125$~mm. Figure~\ref{fig:Fig_S1}(b) shows comparison between experimental results of the microwave absorption parameters \rev{$|S_{11}|$} (dotted green) in a vector network analyser (VNA) and simulated \rev{$|S_{11}|$} (blue) and \rev{$|S_{21}|$} (red) as a function of frequency. Between 1 and 4~GHz, two resonance modes at 1.51 and 2.87~GHz are obtained by simulation, which we respectively attribute the lower and the upper modes as $\omega_{r}\rev{/2\pi }$ and $\omega_{\rm ref}\rev{/2\pi }$. From our experimental \rev{$|S_{11}|$} data (green) with the YIG sample placed on the SRR (see Fig.~1(c) in main text), we find $\omega_{r\rm(exp)}\rev{/2\pi }=1.53$ and $\omega_{\rm ref(exp)}\rev{/2\pi }=3.09$~GHz. Although a trivial shift of $\omega_{\rm ref}$ mode is observed (difference~$\approx7.1~\%$), we notice that the resonance \rev{frequency} of $\omega_{r}$ mode is hardly affected (difference~$\approx1.3~\%$) when the sample is introduced. In Figs.~\ref{fig:Fig_S1}(c)-(e), we respectively show the local distributions for different magnetic field components $H_x$, $H_y$ and $H_z$ of the \rev{$|S_{11}|$} mode at $\omega =\omega_{r}$ (left) and $\omega =\omega_{\rm ref}$ (right). At the resonance frequency, as the microwave currents are flown through the microstrip line, the energy transfer of the electromagnetic fields are absorbed by the SRR, which subsequently induces an oscillating magnetic field excitation in its local geometry.

\section{Spin wave dispersion in the YIG film}

The YIG film used in the experiment is $d=5\ \mu$m thick with magnetic-field ($\mu _0 H$) dependent dispersion relation~\cite{Kalinikos_SPJ1981}
\begin{equation}
    \omega _{n\bm{k}}\left( H\right) = \mu _0 \gamma \sqrt{\left\{ H+M_{\rm s} \left( 1+\lambda _{\rm ex} k_n^2  -P_n \right) \right\} \left\{ H +M_{\rm s} \left( \lambda _{\rm ex}k_n^2 +P_n \sin ^2 \phi \right) \right\} }  .
    \label{eq:Kalinikos}
\end{equation}
Here $M_{\rm s}$, $\gamma $, $\lambda _{\rm ex}$ are the saturation magnetization, gyromagnetic ratio and stiffness constant, respectively. $\bm{k}=(k_x , k_y )^T = k(\cos \phi , \sin \phi )^T$ denotes the wavevector in the plane, $n\pi /d , n\in \mathbb{Z}$ is the quantized out-of-plane component such that $k_n^2 = k^2 +(n\pi /d)^2$, and
\begin{equation}
   P_n = \frac{k^2}{k_n^2}- \frac{2}{1+\delta _{0n}} \frac{k^4}{k_n^4} \frac{1-(-1)^n {\rm e}^{-kd}}{kd} 
\end{equation}
for free surface boundary conditions.

To compute the threshold field value $H_{\rm cr}$ for the occurrence of three-magnon splitting, we study the backward volume wave branch $\phi =0$ without a node in the thickness direction $n=0$, which has the lowest resonance frequency. For each given value of $H$, the energy-momentum conservation requires
\begin{equation}
    H \left( H+M_{\rm s} \right) =4\left( H+ M_{\rm s} \lambda _{\rm ex}k^2 \right) \left\{ H +M_{\rm s} \left( \lambda _{\rm ex} k^2 +\frac{1-{\rm e}^{-kd}}{kd} \right) \right\} . \label{eq:conservation}
\end{equation}
The condition on $k$ for $\omega _{n\bm{k}} \left( H\right) $ to be the bottom of dispersion is
\begin{equation}
    \left( 4\lambda _{\rm ex}k^2 +{\rm e}^{-kd} -\frac{1-{\rm e}^{-kd}}{kd} \right) H + \left( 4\lambda _{\rm ex}k^2 + {\rm e}^{-kd} + \frac{1-{\rm e}^{-kd}}{kd} \right) M_{\rm s} \lambda _{\rm ex}k^2 = 0 . \label{eq:minimum}
\end{equation}
The simultaneous solution of Eqs.~(\ref{eq:conservation}) and (\ref{eq:minimum}) for $(H,k)$ determines $H_{\rm cr}$ and the associated $k_{\rm cr}$ that minimizes $\omega _{n\bm{k}}\left( H_{\rm cr} \right) $. 

\begin{figure}[H]
\centering
\includegraphics[scale=0.6]{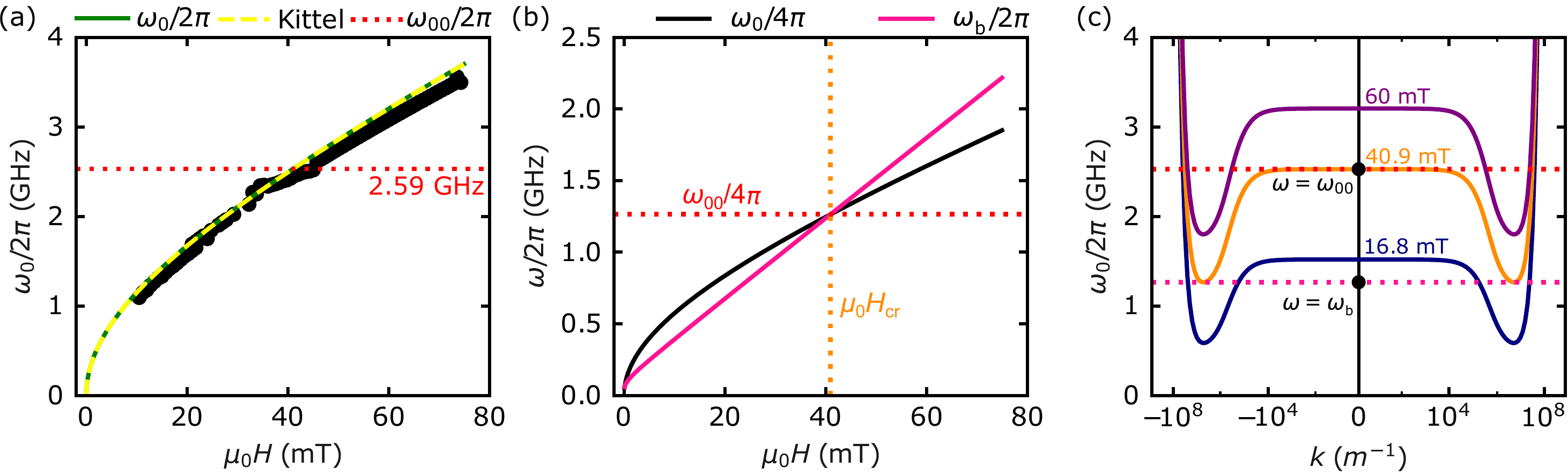}

    \caption{\rev{(a)} Magnetic field dependence of microwave \rev{resonance} frequency $\omega _0/2\pi$. The dots indicate the experimental data in combination with the Kittel fit shown by dashed yellow lines. The solid green curve is the calculated spin wave dispersion obtained by Eq.~(\ref{eq:Kalinikos}) for the lowest energy mode, $n=0$. The dashed red line represents the onset frequency of the three-magnons. \rev{(b)} $\omega_{0}/4\pi$ and $\rev{\omega_{\rm b}/2\pi}$ as a function of magnetic field. The intersection point corresponds to the critical field $\mu_{0}H_{\rm cr}$ for the onset of three-magnon splitting and half the onset frequency. \rev{(c)} Magnon dispersion relationships in YIG of thickness $5\ \mu$m for different external magnetic fields for $\pm \bm{k} \parallel \bm{H}$ calculated using Eq.~(\ref{eq:Kalinikos}).}
    
\label{fig:Fig_S2}
\end{figure}

In order to estimate the value of $M_s$ for our sample, we performed broadband spin wave spectroscopy measurements with our YIG sample placed on a co-planar waveguide (CPW). \rev{$|S_{11}|$} spectra were measured by the VNA for various magnetic fields. We extracted the ferromagnetic resonance (FMR) peak \rev{frequency}~$\omega _{0}\rev{/2\pi }$ from each scan and plot them in Fig.~\ref{fig:Fig_S2}(a). We first fitted the resonance data with the Kittel formula $\omega _{0}=\mu _{0}\gamma \sqrt{H(H+M_{\rm s})}$, and extracted $M_{\rm s} = 1.\rev{26} \times 10^5$~A/m. Note that for $n=0$ and $k\rightarrow 0$, Eq.~(\ref{eq:Kalinikos}) reduces to the Kittel formula.

We now evaluate $H_{\rm cr}$ and $\omega _{0\bm{k}}\left( H_{\rm cr} \right) $. In Fig.~\ref{fig:Fig_S2}(b), we plot half the magnon frequency $\omega_{0}/4\pi$ and the minimum magnon dispersion frequency \rev{$\omega_{\rm b}/2\pi $} as a function of magnetic field, calculated using Eq.~\ref{eq:Kalinikos} with $\gamma =2\pi \times 28$ GHz/T and $\lambda _{\rm ex}=3\times 10^{-16}$ m$^2$~\cite{Stancil_book2008}. The onset condition of three-magnon splitting is satisfied when $\omega_{0}/2 = \rev{\omega_{\rm b} }$, which corresponds to the crossing point in this plot. The numerical solution of Eqs.~(\ref{eq:conservation}) and (\ref{eq:minimum}) yields
\begin{equation}
    k_{\rm cr} = 5.38 \ [{\rm rad/\mu m}], \quad \mu _0 H_{\rm cr} = {\rev{40.9}} \ [{\rm mT}], \quad \frac{\omega _{0\bm{k}}\left( H_{\rm cr} \right)}{2\pi }\bigg| _{\bm{k}\rightarrow \bm{0}} = {\rev{2.59}} \ [{\rm GHz}],
\end{equation}
which gives the coordinate value of the crossing point. The latter value of the critical frequency is quoted in the main text. We identify $\omega _{0\bm{k}}|_{\bm{k} \rightarrow \bm{0}}\rev{/2\pi }$ to be the Kittel mode frequency and denote it by $\omega _0 \rev{/2\pi }$. Finally, we plot in Fig.~\ref{fig:Fig_S2}(c) the spin wave dispersion for different magnetic fields, i.e. the first anticrossing (16.8~mT), $\mu_{0}H_{\rm cr}$, and the second anticrossing (60~mT), to highlight the presence and absence of spin-wave modes with $\rev{\omega _{0\bm{k}}=}\omega_{0}/2$.



    

\section{Data processing methods}
To extract the magnetization properties including the resonance \rev{frequency} ($\omega _0 \rev{/2\pi } $) and the linewidth ($\rev{\kappa _0}/2\pi$), \rev{we initially followed the usual procedure in which} the measured microwave absorption spectra \rev{$|S_{11}|$}~(dB) were first translated into a linear scale to obtain \rev{$\tilde{|S_{11}|}$} (=~$10^{\rev{|S_{11}|}/10}$). Subsequently, the converted data was fit using a double Lorentzian profile:
\begin{equation}
    \rev{\tilde{|S_{11}|}} \propto \frac{A_1\Delta \omega_{1}}{(\omega - \omega_{1})^2+\Delta \omega_{1}^2} + \frac{A_2\Delta \omega_{2}}{(\omega - \omega_{2})^2+\Delta \omega_{2}^2} + C \label{eq:Lorentzian}
\end{equation}
where $\omega$, $\omega_n$, $A_n$, $\Delta \omega_n$, and $C$ correspond to the frequency, resonance position, relative amplitude, linewidth, and the spectra offset, respectively. Here, $n$ denotes the first and the secondary peaks.

\begin{figure}[H]
\centering
\includegraphics[scale=0.6]{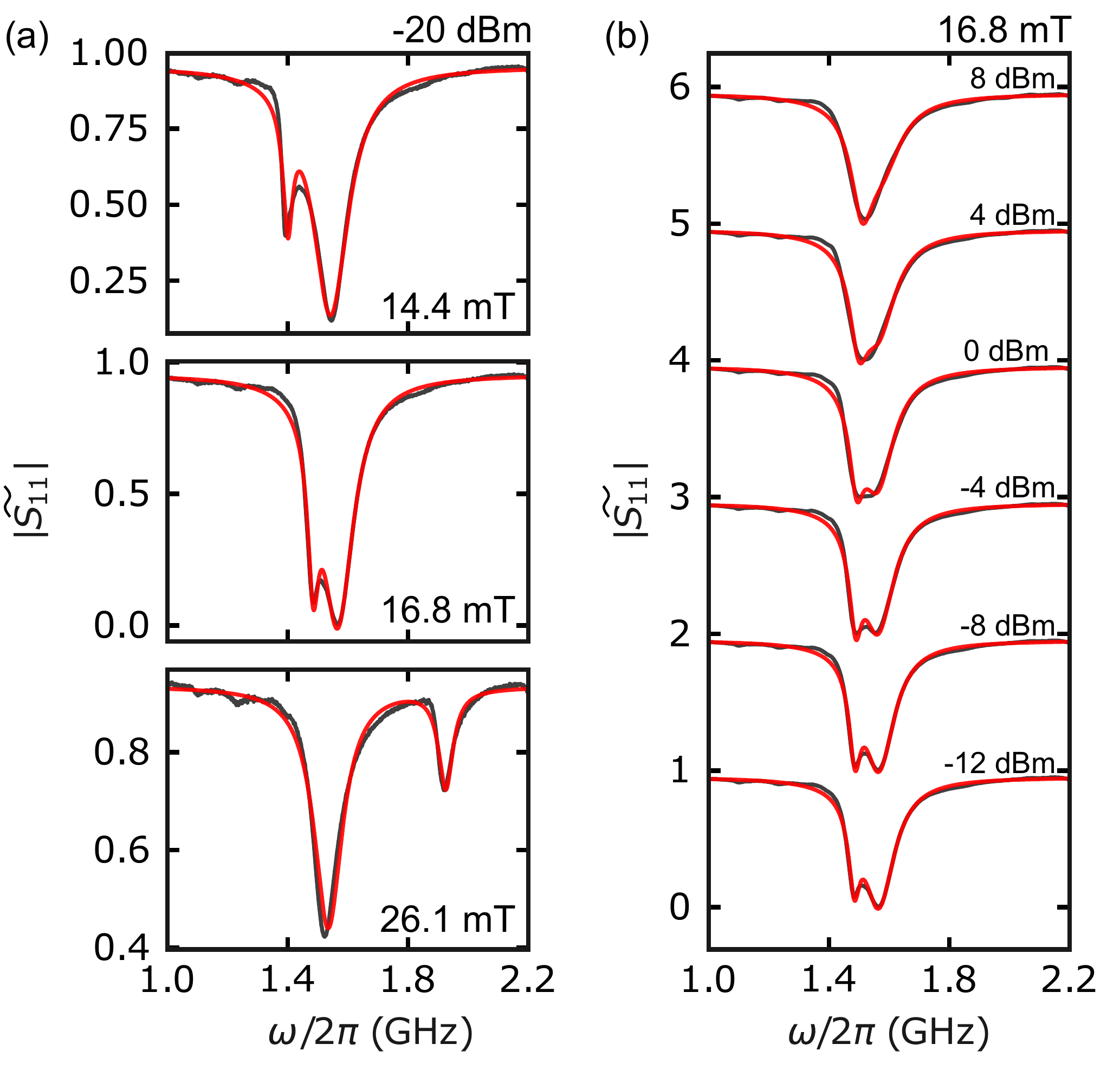}

    \caption{Linear microwave absorption spectra (\rev{$\tilde{|S_{11}|}$}) as a function of frequency (a) at various magnetic fields at a fixed microwave power (-20~dBm) and (b) for different microwave powers at a fixed magnetic field at the \rev{avoided crossing} point (16.8~mT). Red curves are the fits by Eq.~(\ref{eq:Lorentzian}).}
    
\label{fig:Fig_S3}
\end{figure}

In Figs.~\ref{fig:Fig_S3}(a) and (b), we respectively show the individual fitting results of the fields used in the main text at -20~dBm, and for different microwave powers at the \rev{avoided crossing} field (16.8~mT). \rev{As can be seen in Fig.\ref{fig:Fig_S3}(a), the lines look like a well separated double Lorentzian at $\mu _0 H =26.1 $ mT, and the linewidth $\Delta \omega _2 /2\pi$ extracted from this fitting, interpreted as the linewidth~$\kappa _0 /2\pi$ of uncoupled Kittel magnon, has been used in Fig.~2(e) in the main text. When the hybridization is strong at $\mu _0 H = 16.8$ mT, however, the dip between the two peaks is quite shallow even at low input powers. Comparing Fig.~\ref{fig:Fig_S3}(b) with Fig.~2(c) in the main text, one can recognize that the dB unit (by design) is more suitable for quantifying the hybridization gap and its closure at high input powers. Therefore, for the sake of peak frequency determination only, we fitted the dB unit data by the same double Lorentzian (\ref{eq:Lorentzian}) instead of the linearized data and used the result for Fig.~2(d). Furthermore, Fig.~\ref{fig:Fig_S3}(b) indicates that beyond 0 dBm, there is essentially a single peak so that the double Lorentzian fitting is expected to give misleading results. Indeed, we found that the fitting generates values of $\omega _2 /2\pi $ wildly off the main peak frequency $\omega _1 /2\pi \approx 1.52$ GHz. Whenever this happens, we also have $\left| A_2 \right| \ll \left| A_1 \right|$, which implies that the fitting of the second peak does not affect the agreement between the curve and the data points. To avoid confusions, therefore, we removed the experimental data points for $\omega _2$ above 2 dBm in Fig.~4(d).}

\section{Details of the theoretical calculations}

In this section, we describe the theory that leads to Eqs.~(\rev{4}) - (\rev{6}) and Figs.~\rev{2(d), 2(e),} 4(a) and 4(b) in the main text. For convenience of the reader, we repeat the model Hamiltonian
\begin{align}
\mathcal{H} =&  \omega _0 b_0^{\dagger } b_0  + \sum _{\bm{k}\neq \bm{0}} \omega _{\bm{k}}b_{\bm{k}}^{\dagger }b_{\bm{k}}   + \omega _r b_r^{\dagger }b_r + \left[ gb_0^{\dagger }b_r + h{\rm e}^{-i\omega t} \left( U_0 b_0^{\dagger } + U_r  b_r^{\dagger } \right) +{\rm h.c.} \right] \nonumber \\
& +\frac{1}{2}\sum _{\bm{k}\neq \bm{0}}\left( \overline{V}_{\bm{k}}b_0 b_{\bm{k}}^{\dagger } b_{-\bm{k}}^{\dagger } + {\rm h.c.} \right) + \rev{\frac{1}{2}\sum _{\bm{k},\bm{k}^{\prime }} \left( 2T_{\bm{k}\bm{k}^{\prime }} b_{\bm{k}}^{\dagger }b_{\bm{k}^{\prime }}^{\dagger }b_{\bm{k}}b_{\bm{k}^{\prime }} + S_{\bm{k}\bm{k}^{\prime }} b_{\bm{k}}^{\dagger }b_{-\bm{k}}^{\dagger }b_{\bm{k}^{\prime }}b_{-\bm{k}^{\prime }} \right)} . \label{eq:Hamiltonian}
\end{align}
\rev{For brevity, we define $\mathcal{H}$ to have the unit of angular frequency. Throughout this section and the next, ``frequency" and ``magnetic field" refer to angular frequency and $\bm{H}$ instead of $\bm{B} =\mu _0 \bm{H}$ respectively, as is conventional in theoretical literature.} The first line describes the usual linear theory of Kittel mode \rev{magnon} in a microwave cavity augmented by the single particle energy of propagating magnons. The first term in the second line represents the three-magnon interaction arising from the magnetic dipole-dipole interactions, where the coupling constant in a sufficiently thick film is given by
\begin{equation}
V_{\bm{k}} = \omega _M \sqrt{\frac{g_{\rm eff} \mu _B}{2\mathcal{V} M_{\rm s}}} \left( 1+ \frac{\omega _{\bm{k}}}{A_{\bm{k}}+\left| B_{\bm{k}}\right|} \right) \frac{k_x \left( k_y +ik_z \right)}{k^2} , \quad \omega _M = \mu _0 \gamma M_{\rm s} ,  \label{eq:three-magnon}
\end{equation}
with $g_{\rm eff}$ the g-factor, $\mu _B$ the Bohr magneton, $\mathcal{V}$ the volume of the magnet, $k^2 = k_x^2 + k_y^2 + k_z^2 $, and $A_{\bm{k}}, B_{\bm{k}}$ the Bogoliubov coefficients for the magnon eigenstates
\begin{equation}
A_{\bm{k}} = \mu _0 \gamma \left( H  + M_{\rm s} \lambda _{\rm ex} k^2 + M_{\rm s} \frac{k_y^2 +k_z^2}{2k^2} \right) , \quad B_{\bm{k}} = \mu _0 \gamma M_{\rm s} \frac{\left( k_y +ik_z \right) ^2}{2k^2} ,
\end{equation}
with which the eigenfrequency can be written as $\omega _{\bm{k}} =\sqrt{A_{\bm{k}}^2 -B_{\bm{k}}^2}$. This expression could be significantly modified for film thicknesses of a few micrometers, but the detailed form is uncomputable as well as unnecessary. We only assume $V_{-\bm{k}} =V_{\bm{k}}$, which is expected unless the sample shape breaks the inversion symmetry. We set the $x$-axis along the external static magnetic field $H$, the $z$-axis along the film normal, and ignore the crystalline anisotropy so that the ground state magnetization is in the $x$ direction. The Heisenberg Hamiltonian gives rise to many other nonlinear interaction terms between different spin waves, which were all neglected in the main text. Most of them can be removed by a suitable nonlinear transformation of $b_{\bm{k}}$~\cite{Zakharov}. The only \emph{resonant} contributions up to fourth order in $b_{\bm{k}}$, which cannot be dismissed in this way and are relevant under spatially uniform driving, are the latter two terms in the second line in Eq.~(\ref{eq:Hamiltonian}). The expressions for $T_{\bm{k}\bm{k}^{\prime }}$ and $S_{\bm{k}\bm{k}^{\prime }}$ are complicated~\cite{L'vov}, \rev{but their generic order of magnitude estimate is $T_{\bm{k}\bm{k}^{\prime }} \sim S_{\bm{k}\bm{k}^{\prime }} \sim \omega _M / N \ll V_{\bm{k}} \sim \omega _M /\sqrt{N}$ in the GHz range, where $N$ is the number of spins in the sample. Therefore, the three-magnon processes usually dominate, which we assume to be the case. Nevertheless, we do keep $S_{\bm{k}\bm{k}^{\prime }}$, which gives a minor correction to the spectrum, but does turn out to improve the agreement with the experimental data. In contrast, retaining $T_{\bm{k}\bm{k}^{\prime }}$ did not change the result in any significant way and we discard them for simplicity.}

In our convention, the Heisenberg equation of motion reads $dA/dt = i[ \mathcal{H},A]$, which yields
\begin{eqnarray}
i\left( \frac{d}{dt}+\rev{\kappa _r} \right) b_r &=& \omega _r b_r + \overline{g} b_0 + hU_r {\rm e}^{-i\omega t} + \xi _r , \label{eq:photon}  \\
i\left( \frac{d}{dt}+\rev{\kappa _0} \right) b_0 &=& \rev{\left( \omega _0 +S_{00}\left| b_0 \right| ^2 \right)} b_0 + g b_r + h U_0 {\rm e}^{-i\omega t} + \frac{1}{2} \sum _{\bm{k}} \rev{\left( V_{\bm{k}} +2S_{0\bm{k}}\overline{b_0} \right)}  b_{\bm{k}}b_{-\bm{k}} + \xi _0 , \label{eq:Kittel} \\
i\left( \frac{d}{dt} +\eta _{\bm{k}} \right) b_{\bm{k}} &=& \omega _{\bm{k}} b_{\bm{k}} + \left( \overline{V_{\bm{k}}} b_0  + \sum _{\bm{k}^{\prime }} S_{\bm{k}\bm{k}^{\prime }} b_{\bm{k}^{\prime }}b_{-\bm{k}^{\prime }} \right) \overline{b_{-\bm{k}}} + \xi _{\bm{k}} .\label{eq:magnon}
\end{eqnarray}
In writing down \rev{these equations}, we have made a few important alterations to the setup. First of all, the quantum mechanical operators $b_{r,0,\bm{k}}$ have all been relegated to stochastic random variables, and accordingly the Hermitian conjugation $b^\dagger $ has been replaced by the complex conjugation $\overline{b}$. It is justifiable at room temperature where quantum fluctuations are overwhelmed by the thermal fluctuations. Second, we have introduced phenomenologically the relaxation rates \rev{$\kappa _{r,0}, \eta_{\bm{k}}$ that are equal to the frequency linewidth of the respective modes in the linear and uncoupled regime,} and added them to the time derivatives on the left-hand-sides. Corresponding to the relaxation is the random noise fields $\xi _{r,0,\bm{k}}$ that have vanishing average $\langle \xi _{r,0,\bm{k}} \rangle =0$. We assume their higher-order correlations are Gaussian and dictated by the fluctuation-dissipation theorem for $g=V_{\bm{k}} =S_{\bm{k}\bm{k}^{\prime }}=0$ for simplicity:
\begin{equation}
 \left\langle \overline{\xi _{\bm{k}}\left( t\right) } \xi _{\bm{k}^{\prime }}\left( t^{\prime } \right) \right\rangle = 2 \eta _{\bm{k}}   N_{\bm{k}}  \delta _{\bm{k}\bm{k}^{\prime }} \delta \left( t-t^{\prime } \right)  , \quad N_{\bm{k}} =\frac{k_B T}{\hbar \omega _{\bm{k}}} ,
\end{equation}
and similarly for $\langle \overline{\xi _0}\xi _0 \rangle , \langle \overline{\xi _r} \xi _r \rangle $ with all the other correlations being zero. The number of magnons $N_{\bm{k}}$ is assumed to obey Rayleigh-Jeans distribution in the high-temperature approximation. \rev{Finally, we retained only those nonlinear terms that potentially are resonant in the three-magnon processes, eliminating some possible four-magnon terms. It is a working hypothesis in anticipation that the first-order Suhl instability generically precedes the second-order one for $H< H_{\rm cr}$.}

Experimentally, we measure the microwave reflection which is ideally equal to 1 minus the appropriately normalized work per unit time done by the external driving $h$. In order to calculate it, we only need the expectation values of $b_{r,0}$ that oscillate at the driving frequency $\omega $. Therefore, we follow Suhl~\cite{Suhl} and introduce the rotating frame variables
\begin{equation}
c_{r,0} = \left\langle b_{r,0} \right\rangle {\rm e}^{-i\omega t} .
\end{equation}
For the traveling magnons, it turns out that the convenient variables are quadratic correlators
\begin{equation}
s_{\bm{k}} = \left\langle b_{\bm{k}} b_{-\bm{k}} \right\rangle , \quad n_{\bm{k}} = \left\langle \left| b_{\bm{k}} \right| ^2 \right\rangle .
\end{equation}
Foreseeing the subsequent development, let us define the critical amplitude for the Kittel mode by
\begin{equation}
c_{\rm cr} = \frac{1}{V_{\bm{k}}} \left( \omega _{\bm{k}} - \frac{\omega }{2} +i\eta _{\bm{k}} \right) . \label{eq:critical}
\end{equation}
It turns out that for each given $\omega $, the modes with $\bm{k}$ that minimizes $\left| c_{\rm cr} \right| $ become unstable first as the input power (i.e. $\left| h\right|$) is increased. Unless $h$ gets too large beyond a threshold, these remain the only modes that pick up a non-vanishing $s_{\bm{k}}$. This conclusion is reached by studying a zero-temperature formulation which is not a straightforward low-temperature limit of the nonzero-temperature version used here, so that we do not present the details. The reader is referred to Ref.~\cite{L'vov}. We just assume that there is only a single pair $\pm \bm{k}$ of wavevectors that minimizes $\left| c_{\rm cr} \right| $ and drop the summation over $\bm{k}$. This assumption is reasonable when the symmetry of the system is low, but cannot be fully justified due to remaining discrete symmetries, accidental degeneracies, $h$ possibly being significantly above the threshold etc.. We take it as another working hypothesis, supported again by the good agreement with the experimental data. We try and find a stationary state, i.e. a non-trivial time-independent solution for $c_{r,0},s_{\bm{k}},n_{\bm{k}}$. After a straightforward algebra, ignoring the four-magnon contribution, one obtains
\begin{equation}
\left\{ \omega -\omega _0 + i\rev{\kappa _0} -\frac{\left| g\right| ^2}{\omega -\omega _r +i\rev{\kappa _r}} + \frac{V_{\bm{k}} N_{\bm{k}} c_{\rm cr}}{\left| c_{\rm cr} \right| ^2 -\left| c_0 \right| ^2} \right\} c_0 = \left( U_0 + \frac{gU_r}{\omega -\omega _r +i\rev{\kappa _r}} \right) h  , \label{eq:sol_Kittel}
\end{equation}
along with Eqs.~(\rev{5}) and (\rev{6}) in the main text. It further reduces to a real cubic equation for $\left| c_0 \right| ^2$, but the closed-form solution is not very illuminating. We ultimately resorted to numerical solutions \rev{while restoring the relevant four-magnon terms}, which were used to generate Figs. 4(a) and 4(b) in the main text. However, one can guess the qualitative behavior of the solution as a function of $h$ by inspection. For $h\rightarrow 0$, there is clearly only one positive root for $\left| c_0 \right| ^2 \propto \left| h\right| ^2$ that represents the usual FMR solution. It remains a good approximation as long as $\left| c_0 \right| ^2 \ll \left| c_{\rm cr} \right| ^2$. When the latter condition ceases to be valid, the third-term on the left-hand-side starts growing. Note that this term grows indefinitely as $\left| c_0 \right| $ approaches $\left| c_{\rm cr} \right| $ from below. Therefore, however large $\left| h\right| $ becomes on the right-hand-side, there is always a solution $\left| c_0 \right| $ that is close to but no greater than $\left| c_{\rm cr} \right| $. It is easy to convince oneself that this solution continuously evolves out of the usual FMR solution at low power. Thus we are led to conclude that the Kittel mode amplitude saturates at $\left| c_{\rm cr}  \right| $ in the high-power limit. It also tells that at nonzero temperatures, the instability is not a clear threshold process, but it occurs continuously as the input power is increased. The phase information can also be read off from Eq.~(\ref{eq:sol_Kittel}) and Eq.~(\rev{5}) in the main text, which yields Eq.~(\rev{4}) of the main text. \rev{Note that as a function of $\omega $, $c_0$, and therefore $c_r$ and $n_{\bm{k}},s_{\bm{k}}$, cannot be approximated by simple combinations of Lorentzian functions since the denominator $\left| c_{\rm cr} \right| ^2 -\left| c_0 \right| ^2$ also gives a rapidly varying contribution in addition to the usual $\omega -\omega _0 +i\kappa _0$ and $\omega -\omega _r +i\kappa _r$.}

\rev{The discussion above implicitly assumes that the unstable mode has a fixed $\bm{k}$ independent of the input amplitude $\left| h\right|$. It is in general false and the value of $\bm{k}$ continuously evolves as $\left| h\right|$ increases according to a complicated optimization dictated by $\bm{k}$-dependences of $\omega _{\bm{k}}, \eta _{\bm{k}}$, and $V_{\bm{k}}$~\cite{L'vov}. It is practically impossible to be quantitative about all of them, however. Therefore we ignore these $\bm{k}$-dependences, which is expected to capture the qualitative trend. Generally, inclusion of the $\bm{k}$-dependences should slow down the development of instability as a function of $h$ since the nonlinear correction shifts the unstable $\bm{k}$ value from the one that has the optimal condition exactly at the threshold power. In Fig.~2(e), the theory overestimated the nonlinear growth of the linewidth, which may partially be attributed to this simplification in our model.}

We have so far been ambiguous about the driving field amplitude $h$, and it has to be specified for simulations. Here by convention $h$ is taken to have the dimension of frequency, which means $U_{r,0}$ are dimensionless. The threshold value of $h$ for zero temperature is a mess in the general setup. To provide a theoretically convenient normalization, we introduce the threshold field $h_{\rm cr}$ defined for $g=0$:
\begin{equation}
h_{\rm cr} = \frac{\omega -\omega _0 +i\rev{\kappa _0}}{U_0} \frac{\omega /2 -\omega _{\bm{k}}+i\eta _{\bm{k}}}{\overline{V_{\bm{k}}}} .
\end{equation}
It gives an order of magnitude estimate for the relevant input power range, but we do not attach a direct physical meaning to it since the absolute value of $U_0$ is not easily obtainable from the experimental data. The ratio $U_0 /U_r$ affects the lineshape in an essential way, however, \rev{as explained further in the next section}. After experimenting with different values, we fixed $U_0 /U_r = \rev{0.1 +0.3i}$ which gave a reasonable comparison with Fig.~3(c) in the main text. 

To compare the theory with the experimentally measured $\rev{\left| S_{11} \right|} = 10 \log _{10}R $, we need an estimate for the reflection coefficient $R$ that is the reflected power divided by the input power. Since there is no transmission line in our device, the reflected power is equal to the input power $P_{\rm in}$ minus the power dissipated $P_d$ in the device, namely $R= 1-P_d /P_{\rm in}$. By definition $0<R<1$ and therefore $\rev{\left| S_{11}\right| }<0$. The experimental data suggest that $R\approx 1$ away from the resonance and $R\approx 0$ at the peak of the resonance. The former implies that the loss from dynamical entities other than the magnons and photons is very small so that one can reasonably estimate $P_d$ by the dissipation rate of the magnons and photons:
\begin{equation}
P^{\rm theoretical}_d \sim \rev{\kappa _0} \left| c_0 \right| ^2 + \rev{\kappa _r} \left| c_r \right| ^2 + 2\eta _{\bm{k}} n_{\bm{k}} .
\end{equation}
All the quantities on the right-hand-side can be readily computed as functions of $H , \omega $ and $h$ once the model parameters are fixed. To compare the calculation with the experiment, however, we need an expression for $P_{\rm in}$, which is difficult to model theoretically. Here we take a phenomenological approach by observing that in the experimental data the minimum value of $ \rev{\left|S_{11}\right|}$ is roughly -30~dB regardless of the input power. We choose $P_{\rm in}$ for each $h$ such that the computed minimum for \rev{$|S_{11}|$} is equal to -30~dB:
\begin{equation}
S_{11}^{\rm theoretical}= -10 \log _{10} \left( 1 - \frac{\left( 1-10^{-3}\right) P^{\rm theoretical}_d}{\max _{H,\omega } \left[ P^{\rm theoretical}_d \right] } \right) ,
\end{equation}
As explained after Eq.~(\ref{eq:critical}), the wave vector $\bm{k}$ of the magnon pair that become unstable is determined by minimizing $\left| c_{\rm cr} \right| $ for each $\omega $. Therefore, in computing the $\omega $-dependence of $P_d^{\rm theoretical}$, one in principle has to treat $\bm{k}$ as a function of $\omega $. To do it, one would \rev{again} need to know the $\bm{k}$-dependence of $V_{\bm{k}}$ and $\eta _{\bm{k}}$. \rev{As stated above, it is impracticable} and we set both $V_{\bm{k}}$ and $\eta _{\bm{k}}$ as well as $S_{\bm{k}\bm{k}^{\prime }}$ to be independent of $\bm{k}$ and fix their values as
\begin{equation}
V_{\bm{k}} = \omega _M \sqrt{\frac{2g_{\rm eff}\mu _B}{\mathcal{V}M_{\rm s}}} , \quad \eta _{\bm{k}} = \alpha \frac{\omega _0}{2} , \quad \rev{S_{\bm{k}\bm{k}}= S_{0\bm{k}} = 10S_{00} =\omega _M \frac{2g_{\rm eff}\mu _B}{\mathcal{V}M_s}} 
\end{equation}
where the magnon Gilbert damping $\alpha =0.01$ has been introduced and the sample volume is taken to be 
\begin{equation}
\mathcal{V} = \frac{2g_{\rm eff}\mu _B}{M_{\rm s}} \times 10^{11}  .
\end{equation}
Note that the prefactor on the right-hand-side would equal twice the volume of the unit cell if the magnet was a simple ferromagnet. With $V_{\bm{k}}$ and $\eta _{\bm{k}}$ being constants, the minimum of $\left| c_{\rm cr} \right| $ occurs always at $\omega _{\bm{k}} = \omega /2$. For Kittel mode, we decided to use a constant damping $\kappa _0$ independent of $\omega $ instead of the Gilbert type as usually assumed, since the FMR linewidth away from the photon-magnon hybridization appears very weakly dependent of $\omega $, indicating that the disorder-induced two-magnon conbtribution dominates \rev{$\kappa _0$. While the experimental value at $\mu _0 H = 26.1 $~mT is $\kappa _0 = 28$~MHz, it is field-dependent and appears to hit minimum $\kappa _0 =22$~MHz at $\mu _0 H \approx 50$~MHz, which was used in the numerical simulaitons.} Finally, the reference driving field $h_{\rm ref}$ has been taken equal to $h_{\rm cr}$ evaluated at $\omega = \omega _0$ and $\mu _0 H = 15$~mT.

\color{black}
\section{Asymmetry of spectral shapes in the presence of damping}

In the experiment, at low input powers where the system should be described by the linear response theory, we consistently observed that the two peaks of reflection spectrum as a function of $\omega $ for fixed $H$ have different heights even when $H$ is tuned such that the distance between the two peaks is minimized. Further, the peak frequencies do not appear equidistant from the degenerate resonance frequency in the absence of the coupling, which is equal to the resonator eigenfrequency at $H=0$ or $H=\infty $. Here we show that this asymmetry can be attributed to the fact that the system is only marginally in the strong coupling regime and the distortion of the spectral line shapes by the nonzero linewidths cannot be ignored.

Since nonlinearity is irrelevant in the present discussion, we begin with the linearized versions of Eqs.~(\ref{eq:photon}) and (\ref{eq:Kittel});
\begin{equation}
    \begin{pmatrix}
    \omega -\omega _r  +i\rev{\kappa _r} & -\overline{g} \\
    -g & \omega -\omega _0 +i\rev{\kappa _0} \\
    \end{pmatrix} \begin{pmatrix}
    b_r \\
    b_0 \\
    \end{pmatrix} = h \begin{pmatrix}
    U_r \\
    U_0 \\
    \end{pmatrix}
\end{equation}
Solving it for $b_{r,0}$ yields
\begin{eqnarray}
b_r &=& \frac{ \left( \omega -\omega _0 +i\kappa _0 \right) U_r + \overline{g}U_0 }{\left( \omega -\omega _r +i\kappa _r \right) \left( \omega -\omega _0 +i\kappa _0 \right) -\left| g\right| ^2} h , \\
b_0 &=& \frac{\left( \omega -\omega _r +i\kappa _r \right) U_0 +gU_r }{\left( \omega -\omega _r + i\kappa _r \right) \left( \omega -\omega _0 +i\kappa _0 \right) -\left| g\right| ^2} h .
\end{eqnarray}
The total absorbed power should be given by
\begin{equation}
P = \kappa _r \left| b_r \right| ^2 + \kappa _0 \left| b_0 \right| ^2,
\end{equation}
which is the rate of energy dissipation in the stationary state. We are interested in the resonance of the two modes where $\omega _0 = \omega _r $. The power absorption becomes
\begin{eqnarray}
P &=& h^2 \left[ \left\{ \left( \omega -\omega _0 \right) ^2 -\kappa _r \kappa _0 -\left| g\right| ^2 \right\} ^2 + \left( \omega -\omega _0 \right) ^2 \left( \kappa _0 + \kappa _r \right) ^2 \right] ^{-1} \label{eq:power_linear} \\
&& \times \left\{  \left( \kappa _r \left| U_r \right| ^2 + \kappa _0 \left| U_0 \right| ^2 \right) \left( \omega -\omega _0 \right) ^2 + 2\left( \kappa _0 +\kappa _r \right) \Re \left[ gU_r \overline{U_0} \right] \left( \omega -\omega _0 \right) + \left( \kappa _r \kappa _0 +\left| g\right| ^2 \right) \left(  \kappa _0 \left| U_r \right| ^2 + \kappa_r \left| U_0 \right| ^2 \right) \right\} . \nonumber
\end{eqnarray}

Let us first examine how the peak\rev{-frequency} positions are affected by the nonzero linewidths. If $\kappa _0 , \kappa _r \ll \left| g\right|$, the first line in Eq.~(\ref{eq:power_linear}) varies far more rapidly than the second line near the resonance so that the resonance frequencies would be given by
\begin{equation}
    \omega _{\pm } = \omega _0 \pm \sqrt{\left| g\right| ^2 - \frac{\kappa _0^2 + \kappa _r^2}{2}} \approx \omega _0 \pm \left| g\right| . \label{eq:peaks_denominator}
\end{equation}
However, when $\kappa _0 ,\kappa _r $ are comparable to $\left| g\right|$, the $\omega $-dependence of the second line cannot be ignored \emph{a priori}. A necessary condition for a resonance frequency is $dP/d\omega =0$, which can be written as
\begin{eqnarray}
0 &=& \bigg\{ \left( \kappa _r \left| U_r \right| ^2 + \kappa _0 \left| U_0 \right| ^2 \right) \left[ \left( \omega -\omega _0 \right) ^4 -\left( \left| g\right| ^2 +\kappa _r \kappa _0 \right) ^2 \right] \nonumber  \\
&& + \left( \kappa _0 \left| U_r \right| ^2 + \kappa _r \left| U_0 \right| ^2 \right)  \left( \left| g\right| ^2 + \kappa _r \kappa _0 \right) \left[ 2\left\{ \left( \omega -\omega _0 \right) ^2 - \left| g\right| ^2  \right\} +  \kappa _r^2 + \kappa _0^2 \right] \bigg\} \left( \omega -\omega _0 \right) \nonumber \\
&& + \left( \kappa _r + \kappa _0 \right) \Re \left[ gU_r \overline{U_0} \right] \left\{ 3\left( \omega -\omega _0 \right) ^4 - \left( 2\left| g\right| ^2 -\kappa _r^2 -\kappa _0^2 \right) \left( \omega -\omega _0 \right) ^2 -\left( \left| g\right| ^2 +\kappa _r \kappa _0 \right) ^2 \right\} .
\end{eqnarray}
It gives a fifth order algebraic equation in $\omega $ and does not solve analytically. Note that to the leading (first) order in $\kappa _r ,\kappa _0$, the real roots for $\omega $ are $\omega _0 , \omega _0 \pm \left| g\right| $ in agreement with Eq.~(\ref{eq:peaks_denominator}). To identify the perturbation caused by $\kappa _r ,\kappa _0$, let us rewrite it in the dimensionless form as
\begin{equation}
0 = \left[ X^4 - \left( 1 + \epsilon _2 \right) ^2 + 2A \left( 1+ \epsilon _2 \right) \left\{ X^2 -1 + \epsilon _1 \right\} \right] X + B \left\{ 3 X^4 -2 \left( 1-\epsilon _1 \right)  X^2 - \left( 1+\epsilon _2 \right) ^2 \right\} ,
\end{equation}
where
\begin{equation}
    X = \frac{\omega -\omega _0}{\left| g\right|} , \quad  \epsilon _1 = \frac{\kappa _r^2 + \kappa _0^2}{2\left| g\right| ^2} , \quad \epsilon _2 = \frac{\kappa _r \kappa _0}{\left| g\right| ^2} , \quad A = \frac{\kappa _0 \left| U_r \right| ^2 + \kappa _r \left| U_0 \right| ^2}{\kappa _r \left| U_r \right| ^2 + \kappa _0 \left| U_0 \right| ^2} ,\quad B = \frac{\left( \kappa _r + \kappa _0 \right) \Re \left[ gU_r \overline{U_0}\right]}{\left| g\right| \left( \kappa _r \left| U_r \right| ^2 + \kappa _0 \left| U_0 \right| ^2 \right)} .
\end{equation}
If $\epsilon _1 =\epsilon _2 =0$, $X = \pm 1 \Leftrightarrow \omega = \omega _0 \pm \left| g\right| $ solve the equation regardless of the values of $A$ and $B$. However, for our system, $\epsilon _1 \approx 0.91 , \epsilon _2 \approx 0.63$ so that the perturbations are not that small. One can nevertheless see that corrections from $\epsilon _1 $ and $\epsilon _2 $ tend to cancel each other by rewriting
\begin{eqnarray}
0 &=& \left[ X^4 -1 + 2A \left( 1+\epsilon _2 \right) \left( X^2 -1 \right)  +2  \left\{ A \left( 1+\epsilon _2 \right) \epsilon _1 - \epsilon _2 -\frac{\epsilon _2^2}{2} \right\} \right] X \nonumber \\
&& +B \left( 3X^4 -2X^2 -1 + 2\epsilon _1 X^2 -2\epsilon _2 -\epsilon _2^2 \right) .
\end{eqnarray}
As long as the combinations
\begin{equation}
    \epsilon _1^{\prime } = A\left( 1+ \epsilon _2 \right) \epsilon _1 - \epsilon _2 -\frac{\epsilon _2^2}{2} , \quad \epsilon _2^{\prime } = \epsilon _1 - \epsilon _2 - \frac{\epsilon _2^2 }{2}
\end{equation}
remain small, $X=\pm 1$ will be good approximate solutions. For our system, $\epsilon _1^{\prime } \approx -0.11 , \epsilon _2^{\prime }\approx 0.11$, our resonance frequencies are only slightly shifted from $\omega _0 \pm \left| g\right| $. Note that for $B\neq 0$, the corrections to the upper and lower resonance frequencies do not have to be identical, which explains the small asymmetry observed in the experiment.

Finally, let us see that $B\neq 0$ also leads to an asymmetry in peak height. For simplicity, we check the values of $P$ at the unperturbed resonance frequencies $\omega = \omega _0 \pm \left| g\right|$, which yields
\begin{equation}
P \big| _{\omega =\omega _0 \pm \left| g\right|} =  h^2\left( \kappa _r \left| U_r \right| ^2 + \kappa _0 \left| U_0 \right| ^2 \right)  \frac{\left| g\right| ^2 \pm  2B \left| g\right| ^2 + A \left( \left| g\right|^2 + \kappa _r \kappa _0 \right)}{ \kappa _r^2 \kappa _0^2 + \left| g\right| ^2 \left( \kappa _r + \kappa _0 \right) ^2 } .
\end{equation}
It is clear that the peak heights can become asymmetric for $B\neq 0$, which is about 40 percent for $B\approx 0.13$ in our setup (contrast $= 4B / \left\{ 1 + A\left( 1+\epsilon _2 \right) \right\} $), but appears being enhanced by the conversion to dB unit.

\section{Additional data}

\begin{figure}[H]
\centering
\includegraphics[width=0.98\linewidth]{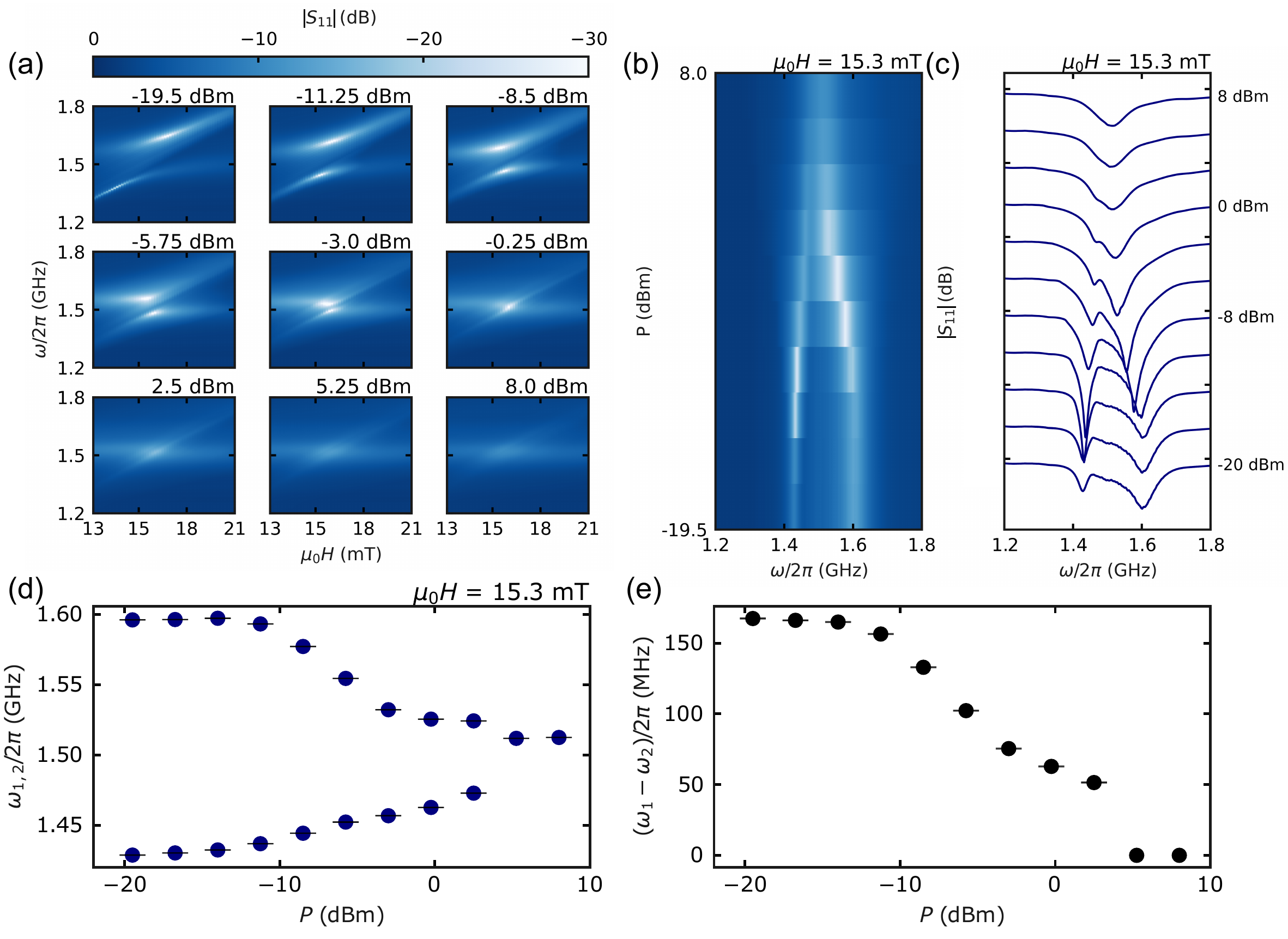}
    \caption{(a) Microwave absorption spectra ($\left| |S_{11}| \right| $ (dB)) as a function of microwave frequency and magnetic field strength for different microwave powers at 100~$\%$ sample coverage placed on the SRR. (b) Collections of frequency domain scans for different microwave powers at a fixed magnetic field, where the mode hybridization is at its minimum. The corresponding individual spectra are shown in (c). (d) Power evolution of extracted peak positions of upper and lower modes ($\omega_{1,2}/2\pi$). (e) The gap size calculated by $\omega_{1,2}/2\pi$.}
    
\label{fig:Fig_S4}
\end{figure}

In this section, we show additional data supporting the observation of nonlinear effects within our experiments. In the first instance, we fully cover the same SRR (in contrast to $\approx~70\%$ coverage in Fig.~2 in main text) as described in Sec. I by the YIG film and conduct power dependence measurements, whose results are shown in Fig.~\ref{fig:Fig_S4}. Here, we take the same analysis method as described for Fig.~2 in the main text. Figure~\ref{fig:Fig_S4}(a) depicts $|S_{11}|$ data for various $P$ as a function of $\omega/2\pi$ and $\mu_{0}H$, while Figs.~\ref{fig:Fig_S4}(b) and \ref{fig:Fig_S4}(c) respectively show the collection and their individual spectra as a function of $P$ at $\mu_{0}H$~=~15.3~mT. Subsequently, we show the extracted peak-frequency positions in Fig.~\ref{fig:Fig_S4}(d) and plot their difference in Fig.~\ref{fig:Fig_S4}(e). In this configuration, we identify that the coupling strength in the low power regime is $\approx$~80~MHz, which is larger than that of Fig.~2 in the main text by a rough factor of 2. While the increased gap size can be explained by the larger mode overlap between the resonator and the YIG film, its closing behavior as a function of $P$ is akin to that of Fig.~2, demonstrating a strong reproducibility of the nonlinear effects.

\color{black}